\documentclass[aps,prl,twocolumn,groupedaddress]{revtex4}

\usepackage{graphicx}
\usepackage{dcolumn}
\usepackage{bm}

\begin{document}

\title{A self-similar ordered structure with a non-crystallographic point symmetry}


\author{Komajiro Niizeki}
\email[]{niizeki.cmpt.phys.tohoku.ac.jp}
\affiliation{Department of Physics, Graduate School of Science, 
Tohoku University, Sendai 980-8578, Japan}

\author{Nobuhisa Fujita}
\affiliation{Department of Structural Chemistry,
Stockholm University,\\ 10691 Stockholm, Sweden}

\date{\today}

\begin{abstract}
A new class of self-similar ordered structures with non-crystallographic point symmetries is presented.
Each of these structures, named superquasicrystals, is given as a section of a higher-dimensional
``crystal" with recursive superlattice structures.
Such structures turn out to be limit-quasiperiodic, distinguishing themselves from quasicrystals which
are quasiperiodic.
There exist a few real materials that seem to be promising candidates for superquasicrystals.

\end{abstract}

\pacs{61.44.Br, 61.50.Ah, 02.20.Hj}

\maketitle




The Penrose patterns reported in 1974, originally introduced as self-similar structures produced by
inflation, are aperiodic structures with long-range positional orders 
as well as non-crystallographic point symmetries \cite{P74, M81}.
Once they were recognized as model structures of quasicrystals (QCs) discovered in 1984 \cite{SBGC}, it was
found that they can be described as quasiperiodic point sets, given as sections of higher-dimensional
crystals, or hypercrystals \cite{LeSt, KaDu, Bak}.
From the viewpoint of aperiodic ordered structures, however,
there can be another type of self-similar structures with long-range positional orders, namely,
limit quasiperiodic structures (LQPSs) \cite{LGJJ, K93, BMS98}. 
In this Letter, we will discuss the following three topics:
i) A limit periodic structure can be described
as a recursive superlattice structure.
ii) An LQPS is given as a section of a {\it supercrystal} 
that is a higher-dimensional limit periodic structure. 
iii) There can exist {\it superquasicrystals} (SQCs), 
which are LQPSs with {\it noncrystallographic point symmetries}. 
Subsequent arguments will start from a brief review of QCs. 
Then a concrete example of octagonal SQCs is introduced, 
and its properties including its structure factor are investigated. 
Finally, the theory is generalized. 

In classical crystallography, crystals are classified into space groups, 
in which point groups and translational groups are combined together. 
Translational symmetry restricts severely point groups for crystals, 
allowing only 2-, 3-, 4-, and 6-fold rotational symmetries. 
Yet noncrystallographic point groups are allowed for QCs including 
${\rm D_8}$ (octagonal), ${\rm D_{10}}$ (decagonal), 
and ${\rm D_{12}}$ (dodecagonal) in 2D and ${\rm Y_h}$ (icosahedral) 
in 3D (for a review for QCs, see \cite{Ya96}). 
A QC having one of these point groups is a $d$-dimensional 
section of a hypercrystal in $2d$-dimensions, with $d=2$ or $3$ 
being the number of physical space dimensions. 
The entities of the hypercrystal are not atoms but geometric objects 
called hyperatoms (atomic surfaces, windows, acceptance domains, etc.). 
We consider a hypercrystal constructed by locating one kind of hyperatoms 
onto the sites of a Bravais lattice ${\hat \Lambda}$ in the $2d$-dimensional space 
$E_{2d}$. 
The point group ${\hat G}$ of ${\hat \Lambda}$ is isomorphic 
with a noncrystallographic point group $G$ in $d$ dimensions. 
More precisely, ${\hat G}$ is a direct sum of two groups $G$ and $G^\perp$
that act on the physical space $E_d$ and the orthogonal complement
$E_d^\perp$, respectively, where all the three groups ${\hat G}$, $G$,
and $G^\perp$ are isomorphic with each other.

We now begin with some preliminary investigation of an octagonal QC in 2D,
which will set the basis of introducing SQCs in a later paragraph.
For the octagonal QC, the relevant lattice ${\hat \Lambda}$ is a hypercubic
lattice in 4D. 
Let $\Lambda$ be its projection onto the physical space $E_2$. 
Then, it is an additve group composed of vectors such that 
the addition of its two members belongs to it and any member, $\ell$, 
and its inversion, $-\ell$, belong together to it; $\Lambda$ is mathematically a module. 
Every ``lattice vector" $\ell$ in $\Lambda$ is given uniquely as an integral
linear combination of the generators, ${\bf e}_i$ ($i = 0 - 3$), which are 
the projections of the four standard basis vectors of ${\hat \Lambda}$ onto $E_2$.
These generators point toward four successive vertices of a regular
octagon centered at the origin, and the point symmetry of $\Lambda$ is equal
to ${\rm D_8}$. 
$\Lambda$ is a dense set in $E_2$ because the number of its generators is more than two. 
If one defines $\Lambda^\perp$ and ${\bf e}_i^\perp$ on $E^\perp$ in a
similar way, it can be shown that ${\bf e}_i^\perp = (-1)^i{\bf e}_i$, 
which means that $\Lambda$ and $\Lambda^\perp$ are identical as sets of vectors.
Obviously the 4D vectors ${\hat {\bf e}}_i := ({\bf e}_i, \,{\bf e}_i^\perp)$ are the basis
vectors of ${\hat \Lambda}$, and the pair of a lattice vector $\ell$ in
$\Lambda$ and its conjugate $\ell^\perp$ in $\Lambda^\perp$ form a 4D
lattice vector ${\hat \ell} := (\ell, \,\ell^\perp)$ in ${\hat \Lambda}$. 
If the hyperatom is a regular octagon such that its sides are
translates of the eight vectors, $\pm{\bf e}_i^\perp$, the resulting
octagonal QC is a discrete subset of $\Lambda$, yielding an octagonal tiling
with a square and a 45$^\circ$ rhombus; 
the ``bonds" of the tiling are translates of $\pm{\bf e}_i$ \cite{Ya96}. 

\begin{figure}[hbt]
\begin{center}\includegraphics[width = 4cm]{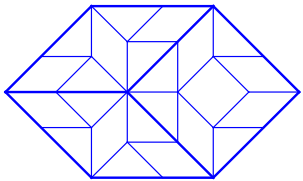}
\caption{
Inflation rule of the octagonal QC (tiling): The inflated tiles are sided by thicker lines. 
The rhombic tile after the inflation is symmetrically decorated with the original tiles 
but the square one is asymmetrically. 
The square has a polarity along a diagonal and the polarity is determined 
by the presence of twinned rhombi sharing sides with the square. 
}
\label{Fig.1}
\end{center}\end{figure}

We have already seen several examples of triplets of objects, 
$\{X, \, X^\perp, \,{\hat X}\}$, associated with the three worlds, $E_2$, $E_2^\perp$
and $E_4$, where the latter two of the triplets are uniquely determined
by the first, $X$ \cite{KaDu}. 
There exists an important triplet of linear maps
$\{\varphi, \, \varphi^\perp, \,{\hat \varphi}\}$ with $\varphi$ (resp. $\varphi^\perp$)
being a scaling transformation with the ratio $\tau :=1 + \sqrt{2\,}$
(resp. ${\bar \tau} :=1 - \sqrt{2\,}$) whereas ${\hat \varphi}$ is a 4D map defined as 
${\hat \varphi} = \varphi \oplus \varphi^\perp$ or, equivalently,
${\hat \varphi} = \left ({\varphi \atop 0}\; {0 \atop \varphi^\perp}\right)$; 
we may call ${\hat \varphi}$ a bi-scaling. 
Note that ${\bar \tau}$ is the algebraic conjugate of $\tau$ and written as ${\bar \tau} = -1/\tau)$. 
A simple geometrical consideration proves that $\tau{\bf e}_i$ 
and $\tau^{-1}{\bf e}_i$ belong both to $\Lambda$, 
and $\Lambda$ has the scaling symmetry,
$\tau\Lambda = \Lambda$, where $\tau\Lambda$ stands for the set of all
the vectors of the form $\tau\ell$ with $\ell$ in $\Lambda$. 
It follows that ${\hat \varphi}{\hat \Lambda} = {\hat \Lambda}$. 
In fact, we may write ${\hat \varphi}({\hat {\bf e}}_0\,{\hat {\bf e}}_1\,{\hat {\bf e}}_2{\,\hat {\bf e}}_3) = 
({\hat {\bf e}}_0\,{\hat {\bf e}}_1\,{\hat {\bf e}}_2{\,\hat {\bf e}}_3)M$ with 
\begin{eqnarray}
M = \left(\matrix{1 & 1 & 0 & -1 \cr 1 & 1 & 1 & 0 \cr 0 & 1 & 1 & 1 \cr -1 & 0 & 1 & 1 \cr}\right)
\end{eqnarray}
being a unimodular 4D matrix \cite{Ni89}: $\det M = 1$. 
We shall call $M$ the companion matrix of ${\hat \varphi}$. 
Using the fact that $\varphi$ is expansive and $\varphi^\perp$ contractive, 
we can show that the octagonal QC above has a self-similarity with ratio
$\tau$ \cite{Ni89}. 
The self-similarity is represented as the inflation rule for the two types
of tiles (see Fig.1). 
Note that a recursive structure is built in the QC by its self-similarity. 

Now let us divide ${\hat \Lambda}$ into two sublattices with respect to
the parity of the sum of indices of the lattice vectors.
The two sublattices are equivalent, and the volume of their unit cells is
twice that of ${\hat \Lambda}$ \cite{Ni89}; the multiplicity of each sublattice is equal to 2. 
Every ``bond" of the octagonal tiling connects two vertices with opposite parities. 
The vertices of the inflated tiling come evenly from the two equivalent
sublattices because the scaling $\varphi$ does not change the parity of a lattice vector. 
For the sake of a later argument, 
we shall investigate the even sublattice, ${\hat \Lambda}_{\rm e}$, of ${\hat \Lambda}$. 
It is a simple algebra to show that ${\bf e}'_i := {\bf e}_i + {\bf e}_{i+1}$
with ${\bf e}_4 := -{\bf e}_0$ are generators of $\Lambda_{\rm e}$, i.e., the projection of ${\hat \Lambda}_{\rm e}$. 
We may write ${\bf e}'_i := \sigma{\bf e}_i$ and, hence, $\Lambda_{\rm e} = \sigma\Lambda$, 
where $\sigma$ is a similarity transformation 
that is a combined operation of a scaling through 
$2\cos{(\pi/8)} \,(= (2 + \sqrt{2\,})^{1/2})$
and a rotation through 22.5$^\circ$ ($= \pi/8$). 
The conjugate $\sigma^\perp$ of $\sigma$ is a scaling through 
$2\sin{(\pi/8)} \,(= (2 - \sqrt{2\,})^{1/2} \approx 0.765)$
and a rotation through -157.5$^\circ$ ($= -7\pi/8$). 
The 4D lattice ${\hat \Lambda}_{\rm e}$ is written with 
${\hat \sigma} = \sigma \oplus \sigma^\perp$ as ${\hat \Lambda}_{\rm e} = {\hat \sigma}{\hat \Lambda}$; 
we may call ${\hat \sigma}$ a bi-similarity transformation. 
The companion matrix of ${\hat \sigma}$ is given by 
\begin{eqnarray}
\left(\matrix{1 & 0 & 0 & -1 \cr 1 & 1 & 0 & 0 \cr 0 & 1 & 1 & 0 \cr 0 & 0 & 1 & 1 \cr}\right), 
\end{eqnarray}
which is not unimodular: $\det M = 2$. 
We may consider ${\hat \Lambda}_{\rm e}$ to be a superlattice of ${\hat \Lambda}$. 
An infinite series of modules is generated as 
$\Lambda_n := \sigma^n\Lambda$, $n = 1, \, 2, \,\cdots$, 
each of which is a projection of the relevant superlattice 
${\hat \Lambda}_n \;(= {\hat \sigma}^n{\hat \Lambda})$; 
the index of $\Lambda_n$ in $\Lambda$ is equal to $2^n$. 
Note that $\Lambda_2 = (2 + \sqrt{2\,})\Lambda = \sqrt{2\,}\Lambda$ 
because $\Lambda$ is invariant against both the 45$^\circ$ rotation and the $\tau$-scaling. 

\begin{figure}[hbt]
\begin{center}\includegraphics[width = 8.5cm]{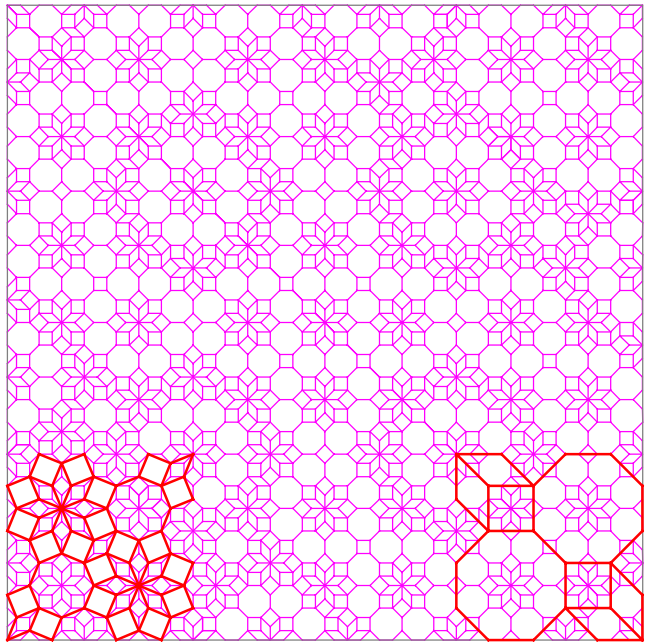} 
\caption{A self-similar octagonal tiling in 2D: 
It is produced by an inflation rule for the three kinds of tiles. 
The left-bottom corner is the center of the exact octagonal symmetry. 
This center is the origin of $E_2$, so that it is an even site. 
A part of the inflated tiling is shown at the left-bottom, 
while a part of the doubly inflated one is at the right-bottom. 
} \label{Fig.2}
\end{center}\end{figure}

We show in Fig.2 another self-similar octagonal tiling produced by an inflation rule. 
It is shown later on that the vertices of the tiling form an SQC.
Since the ``bonds" of the tiling are translates of $\pm{\bf e}_i$ 
the SQC is a discrete subset of $\Lambda$, and is given as a section of 
some 4D structure on ${\hat \Lambda}$; we shall call the 4D structure a supercrystal. 
Inspecting Fig.2 we find that the vertices of the inflated SQC come only from 
the even sublattice, ${\hat \Lambda}_{\rm e}$, and 
the inflation is equivalent to a similarity transformation through $\sigma$.
Moreover, the double inflation of the SQC in Fig.2 is equivalent to 
the $2 + \sqrt{2\,}$-scaling of the SQC. 
It is now evident that the recursive structure being built in the SQC by its self-similarity 
requires that the supercrystal has a {\it recursive superlattice structure} 
associated with the infinite series of superlattices. 
Therefore, the supercrystal is formed by hyperatoms 
whose shapes, sizes, and/or orientations depend on the local environments 
of the relevant sites of the supercrystal; the hyperatoms are {\it not uniform}. 

\begin{figure}[hbt]
\begin{center}\includegraphics[width = 8.5cm]{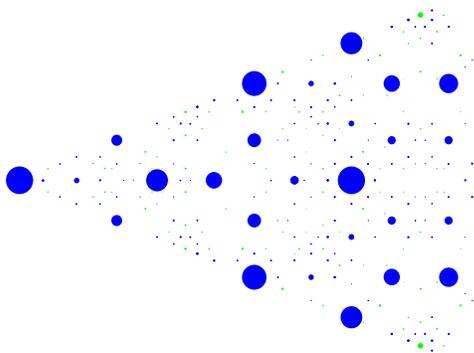}
\caption{
A 1/8 sector of the structure factor of the octagonal SQC: 
Reflections with lower intensities than 0.001 in unit of the maximal value are ignored. 
The blue spots show main reflections, 
while those in green and red show the first-order superlattice 
reflections and the second-order ones, respectively. 
The area of a spot is proportional to the relevant intensity. 
} \label{Fig.3}
\end{center}\end{figure}

The structure factor of the tiling in Fig.2 is composed only of Bragg
spots as shown in Fig.3, so that the tiling has a long-range order.
Their position vectors form the Fourier module, 
which corresponds to the reciprocal lattice for a periodic crystal.
Strong Bragg spots appear at common positions to those of the octagonal QC shown in Fig.1.
They belong to a module given as the projection of the reciprocal lattice 
of ${\hat \Lambda}$ onto $E_2$, and are indexed by four integers \cite{Ya96}.
The structure factor of the SQC, however, includes extra reflections 
originating from members of the infinite series of superlattices described above, 
and {\it the Bragg spots are indexed generally by five integers}; 
the fifth integer specifies the superlattice order $n$ with $n = 0$ for main reflections \cite{Genr}. 
This proves that the SQC is limit quasiperiodic, 
and the supercrystal is not periodic but limit periodic \cite{LGJJ, K93, BMS98}. 
The present SQC is a 2D section of a 4D {\it supercrystal} 
constructed on a {\it recursive superlattice}. 
The hyperatoms are, actually, not so non-uniform for the SQC in Fig.2 
because all the superlattice reflections are weak.
However, different choices of hyperatoms are allowed on the same recursive superlattice, 
yielding different SQCs; some of them can have strong superlattice reflections. 
Each of these SQCs nevertheless has its own self-similarity whose ratio is
written as $2\cos{(\pi/8)}\tau^k$ with $k$ being a nonnegative integer (cf. \cite{Ni89}).
The Fourier module is common among them, so that the recursive superlattice
is considered to be a sort of Bravais lattice. 

While a QC or an SQC is defined as the projection of a subset of ${\hat \Lambda}$ onto $E_2$, 
its conjugate set is defined as the projection of the same set onto $E_2^\perp$. 
The conjugate set of the octagonal QC is a dense set bounded by the same octagon 
as the hyperatom. Hence the shape of the hyperatom is retrieved from the conjugate set. 
The conjugate set of the octagonal SQC is not so simple because hyperatoms are not uniform. 
If we ignore higher order superlattice structures than a specified order, $n$, 
we obtain a hypercrystal whose Bravais hyperlattice is equal to ${\hat \Lambda}_n$; 
its unit cell includes distinct hyperatoms. It yields the $n$-th {\it approximant QC} to the SQC. 
Our investigation of hyperatoms of succesive approximants strongly indicates 
that the hyperatoms of the present SQC are topologically disks 
but their boundaries are fractals. 
Nevertheless the present SQC satisfies the so-called gluing 
(or closeness) condition (for the gluing condition, see \cite{Ka89}).

The two maps, $\varphi$ and $\sigma$, are {\it Pisot maps}; a general Pisot map $\psi$  
is a member of the triplet $\{\psi, \, \psi^\perp, \,{\hat \psi}\}$ such that 
i) ${\hat \psi}$ is a bi-similarity transformation. 
ii) $\psi$ is expansive but $\psi^\perp$ contractive. 
iii) $\psi\Lambda$ is identical to or a submodule of $\Lambda$. 
Conversely, any submodule of $\Lambda$ can be given as $\psi\Lambda$
with a Pisot map $\psi$ provided that the submodule is geometrically similar to $\Lambda$ \cite{Ni89}; 
the index of the submodule in $\Lambda$ is given 
by $m := |\!\det M|$ with $M$ being the companion matrix of ${\hat \psi}$. 
The structure factors of self-similar structures based on Pisot maps 
are known to be composed only of Bragg spots \cite{LGJJ,K93}. 
The companion matrix of the Pisot map associated with a QC is ``unimodular" because $m = 1$, 
while the one with an SQC is not because $m \ge 2$. 
A recursive superlattice is considered to be a Bravais hyperlattice, 
and an infinite number of SQCs are constructed on it. 
Then, it is important to classify possible recursive superlattices with 
non-crystallographic point symmetries that are physically important. 
A recursive superlattice is constructed on the {\it base lattice} ${\hat \Lambda}$ 
by the use of a Pisot map $\psi$. 

We begin with the case of octagonal SQCs in 2D. 
The module $\Lambda$ and its submodule $\psi\Lambda$ have the point group ${\rm D_8}$ 
as their common symmetry group if and only if 
$\psi$ is a simple scaling transformation or a similarity transformation 
including a rotation through 22.5$^\circ$ ($= \pi/8$) \cite{Ni89}; 
a Pisot map of the former (resp. latter) type shall be classified into the type I (resp. II).
The type I Pisot map is a bi-scaling, ${\hat \rho}$, specified by so-called a Pisot number, 
which is a positive number of the form $\nu : = p + q\tau$ with $p, \,q \in {\bf Z}$ and 
the magnitude of its algebraic conjugate, ${\bar \nu}$, is smaller than one. 
It can be readily shown that $\rho = p1 + q\varphi$, ${\hat \rho} = p{\hat 1} + q{\hat \varphi}$ 
and $M_{\hat \rho} = pI + qM_{\hat \varphi}$. 
It follows that $m = [N(\nu)]^2$ with $N(\nu) := \nu{\bar \nu} \;(= p^2 +2pq - q^2)$. 
On the other hand, the type II Pisot map is written as $\psi = \rho\sigma$, 
so that ${\hat \psi} = {\hat \rho}{\hat \sigma}$ and $M_{\hat \psi} = M_{\hat \rho}M_{\hat \sigma}$. 
It follows that $m = 2[N(\nu)]^2$. 
Thus an infinite number of recursive superlattices with the octagonal symmetry can be 
constructed on the octagonal hyperlattice \cite{Ni89}, 
whereas there is only one hyperlattice (exactly, Bravais hyperlattice) 
with the octagonal point symmetry \cite{Ya96}.

The above consideration for the octagonal SQCs is readily extended to the decagonal case (${\rm D_{10}}$), 
the dodecagonal case (${\rm D_{12}}$), and the icosahedral case (${\rm Y_h}$); 
the last case is for 3D SQCs. 
We have succeeded in a complete classification of recursive superlattices with these point symmetries. 
A simplest decagonal SQC, for example, is specified by a type II  Pisot map, and 
its  multiplicity (index) is five, while a simplest icosahedral SQC is specified by a type I  Pisot map, and 
its  multiplicity (index) is 64. 
A complete list will be published elsewhere. 

The selfsimilarity of a QC or an SQC is based on a Pisot map associated with the hyperlattice ${\hat L}$. 
The presence of a Pisot map is derived from the $(d + d)$-reducibility of 
the point group ${\hat G}$ of the hyperlattice ${\hat L}$ \cite{KaDu, Ni89}. 
The $(d + d)$-reducibility is, in turn, a consequence of the fact that the non-crystallographic point group $G$ 
is lifted up to ${\hat G}$. 
Thus, the selfsimilarity of a QC or an SQC is closely connected with its non-crystallographic point symmetry. 
A QC and an SQC are distinguished by whether the relevant Pisot map is volume-conserving or not, respectively. 
QCs and SQCs form an important class of aperiodic ordered structures with non-crystallographic point symmetries, 
and {\it the QCs form a special sub-class which, albeit important, is a definite minority in the class}. 
One may consider an SQC as a bizarre object because of its limit-quasiperiodic nature 
but it is as a natural object as a QC is as mentioned above. 

A 2D QC has a set of parallel quasi-lattice-lines \cite{KaDu} 
so that all its lattice points are located dividedly on them. 
The same is true for an SQC; the quasi-lattice-lines form just a limit quasiperiodic grid, 
whose spacings take the form $p + q\tau$ with $p, \,q \in {\bf Z}$ 
in an appropriate length unit. 
This allows us to relate 2D SQCs to the 1D analogues, which have
been extensively investigated \cite{LGJJ, K93}. 
It is readily shown that SQCs formed by this method satisfy the gluing condition. 
Moreover, the hyperatoms are shown to be polygons, so that their boundaries are not fractals. 
The grid method is extended to the 3D case. 

An exact mathematical formulation of the limit periodic structure is made by using 
locally-compact Abelian groups \cite{BMS98}, 
which are rather transcendental objects for physicists in 
condensed matter physics or for crystallographers. 
We have succeeded in reformulating it on the basis of the recursive superlattice structure, 
which is basically an ``inverse system" \cite{BMS98}. 
We will not, however, present these technical details here; instead,
we enumerate several merits of this approach.
i) This approach to SQCs is a simple extension of the super-space approach to QCs \cite{Ya96}. 
ii) It explains naturally why the structure factors of SQCs are composed only of Bragg spots. 
iii) The Fourier modules of SQCs are readily calculated (cf. \cite{LGJJ, K93}). 
iv) Non-crystallographic point symmetries are easily included. 

Quite recently, the MLD (Mutual-Local-Derivability) classification 
of quasicrystals has been completed \cite{Ni04}. 
It is an urgent task to perform an MLD classification of SQCs. 
Prior to do it, however, one must specify all the SQCs satisfying the gluing condition. 

A superlattice ordering associated with the sublattice with multiplicity (index) five was observed in an
Al-Cu-Co decagonal quasicrystal \cite{EST}. 
An Al-Pd-Mn icosahedral quasicrystal is known, on the other hand, to exhibit a superlattice
ordering such that its hyperlattice constant is doubled \cite{IM92}. 
A superlattice ordering with a large multiplicity as this case is unusual for
conventional alloys but could be interpreted naturally as an icosahedral SQC. 
Anyway, these two ``QCs" are promising candidates for SQCs. 

SQCs as well as QCs are quite natural structures from the point of view of
the generalized crystallography. 
It is surprising that such an important class of structures as that of
SQCs has been entirely overlooked up to the present. 
Discovering and/or synthesizing SQCs is a big challenge in material science, 
and its achievement will be a triumph of the generalized crystallography. 
Moreover, there exists strong evidence that the nature of the electronic states in 
an SQC is markedly different from that in a QC \cite{ENF}. 

\bigskip
\noindent {\bf Acknowledgements}

\noindent {\small The authors are grateful to O. Terasaki.


\begin{thebibliography}{50}
\bibitem{Genr} This does not mean, however, that the Fourier module of the SQC 
is generated by five generators because a structure having a Fourier module 
generated by finite bases must be periodic or quasiperiodic \cite{Ya96}.
\end{thebibliography}
\end{document}